\documentclass[a4paper,fleqn]{cas-sc}

\usepackage[authoryear]{natbib}

\def\tsc#1{\csdef{#1}{\textsc{\lowercase{#1}}\xspace}}
\tsc{WGM}
\tsc{QE}
\tsc{EP}
\tsc{PMS}
\tsc{BEC}
\tsc{DE}

\begin{document}
\let\WriteBookmarks\relax
\def\floatpagepagefraction{1}
\def\textpagefraction{.001}
\shorttitle{Enriching User Shopping History: Empowering E-commerce with a Hierarchical Recommendation System}
\shortauthors{Irem Islek et~al.}

\title [mode = title]{Enriching User Shopping History: Empowering E-commerce with a Hierarchical Recommendation System}




\author[1, 3]{Irem Islek}
\cortext[mycorrespondingauthor]{Corresponding author}
\ead{isleki@itu.edu.tr}
\author[2, 3]{Sule Gunduz Oguducu}
\address[1]{Department of Computer Engineering, Istanbul Technical University, Istanbul, Turkey}
\address[2]{Department of Artificial Intelligence and Data Engineering, Istanbul Technical University, Istanbul, Turkey}
\address[3]{ITU AI Research and Application Center, Istanbul, Turkey}

\begin{abstract}
Recommendation systems can provide accurate recommendations by analyzing user shopping history. A richer user history results in more accurate recommendations. However, in real applications, users prefer e-commerce platforms where the item they seek is at the lowest price. In other words, most users shop from multiple e-commerce platforms simultaneously; different parts of the user's shopping history are shared between different e-commerce platforms. Consequently, we assume in this study that any e-commerce platform has a complete record of the user's history but can only access some parts of it. If a recommendation system is able to predict the missing parts first and enrich the user's shopping history properly, it will be possible to recommend the next item more accurately. Our recommendation system leverages user shopping history to improve prediction accuracy. The proposed approach shows significant improvements in both NDCG@10 and HR@10.\end{abstract}



\begin{keywords}
User History Enrichment \sep Recommendation Systems \sep Transformers 
\end{keywords}

\maketitle

\section{Introduction\label{introduction}}
The recommendation system, a key element of an e-commerce platform, provides many benefits both to the user and the e-commerce platform. From the perspective of the user, it can take quite a long time to search for the item that suits the user's interest and taste among the wide item range in an e-commerce platform without support and help from the recommendation system. On the other hand, from the perspective of the e-commerce platform, recommendation systems directly increase profits due to their positive direct effects on user engagement. In light of all this information, recommendation systems are becoming increasingly popular and influential. In particular, identifying and solving various domain-specific problems is one of the development areas of recommendation systems.
For the e-commerce domain, one of the most crucial needs is to evaluate the user's shopping history properly and develop recommendation systems capable of making successful recommendations even from the shortest shopping histories available.

In the literature, there are different approaches to recommendation systems that can be divided into two categories: traditional and deep learning-based methods. 

According to Adomavicius \citep{AdomaviciusSurvey}, traditional recommendation systems can be divided into three main categories: Collaborative Filtering, Content-Based, and Hybrid Recommendation Models. Collaborative Filtering studies user-item interactions to recommend the next item, while Content-Based studies use content information to make recommendations. Hybrid recommendation studies combine the advantages of both models to develop more effective recommendation systems.

Deep learning-based approaches have been frequently used in recent years because they are more successful in modeling the relationship between items in the users’ shopping history. Deep learning-based recommendation systems have achieved great success using various approaches such as Multi-Layer Perceptron (MLP) \citep{MLP1, MLP2, MLP3}, DeepFM \citep{DeepFM}, Wide \& Deep Learning \citep{Wide&Deep}, and sequential modeling based on Recurrent Neural Networks (RNN) \citep{GRU3, GRU4, Gru4Rec} and Long-Short Term Memories (LSTM) \citep{LSTM1, LSTM2}. In addition, attention models have proven effective in sequential modeling, particularly in the Natural Language Processing (NLP) domain, and have therefore been applied to recommendation systems as well \citep{Attention2, Attention4}. One of the most successful self-attention-based recommendation systems is SASRec \citep{SasRec}, which outperforms previous studies. Other studies utilize BERT \citep{Bert} in content encoding thanks to its capabilities in the textual content.

Despite the improvements provided by the above studies, the problem of giving accurate recommendations to users whom we know little about them is still not considered completely resolved for the e-commerce domain. To solve this problem, we look at the issue from the following perspective: In the real world, generally, users prefer e-commerce platforms where the item they want is the lowest price. When users want to buy an item, they usually look at several e-commerce platforms for that item and prefer to shop from different e-commerce platforms for different reasons. In other words, most users shop from multiple e-commerce platforms simultaneously; different parts of the user's shopping history are shared between different e-commerce platforms. Consequently, this study assumes that no e-commerce platform has a complete record of the user's shopping history but only accesses some parts. In this study, we focus on designing a hierarchical recommendation system for an e-commerce platform that fills users' shopping history gaps by training a bidirectional encoder representation model. Enriching users' shopping history allows us to predict the next item a user will likely interact with more accurately.

In order to make it more clear, we would like to explain more about what we mean by enriching the user shopping history. Various positions selected within the user shopping history are considered places where the user has purchased one or more items from another e-commerce platform. The proposed approach predicts the item the user may have purchased for a given position using the previous and subsequent items. Then, it recommends the following item that the user will be interested in based on this enriched shopping history. 


We think that the hierarchical approach we propose has two significant advantages: (1) it will be possible to make accurate recommendations about users with whom we have little knowledge, thanks to the enriched user-item shopping history. (2) It will be possible to make much more accurate recommendations by obtaining a much richer user-item shopping history, even for users we think we have enough information about them. 

The recommendation system we propose has a hierarchical structure. This hierarchical structure consists of two layers: (1) The first layer takes the user-item shopping history, which includes the user's past shopping information, as an input and outputs an enriched user shopping history with the additions made to various intermediate locations in this history. (2) The second layer takes the enriched user shopping history of the user as input and recommends the following item the user will be interested in as output.

We evaluated the hierarchical model proposed in \citep{ni2019justifying} using three real-world datasets obtained. Enriching the user's shopping history improves item recommendation performance.

We aim to answer the following research questions with this study:
\begin{itemize}
\item \textbf{RQ1:} Does enhancing the user's shopping history improve the recommendation model's performance?

\item \textbf{RQ2:} Which of the different approaches to determining which location to choose for enrichment in the user's shopping history provides more precise user shopping history enrichment?

\item \textbf{RQ3:} When enriching the user's shopping history, does the performance of recommending the next item increase as we continue to enrich?
\end{itemize} 

Our contributions are as follows:
\begin{itemize} 
\item In recommendation models developed for e-commerce, it is crucial to have more historical information about the user to increase the performance of predicting the next item. The hierarchical approach we propose ensures that it first enriches the user's shopping history and recommends the next item to that user using their enriched shopping history. In this way, our approach increases the performance of recommending the next item. We have addressed our first research question with this contribution.

\item The user shopping history enrichment model, which is the first step of the hierarchical approach we propose, provides benefits not only for users with little shopping history but also for users about whom we have shopping history with sufficient length. This contribution is also related to the first research question.

\item The approach we propose for user shopping enrichment predicts an item for an imaginary mask position which we add to the user shopping history using bidirectional encoder representations. Thus, the user shopping history is enriched thanks to the bidirectional encoder-based model.

\item For user shopping enrichment, we determined the right approach to choose where an item likely to be bought by the user from another e-commerce platform will be positioned within the user's shopping history. We tested which positions will be enriched with different approaches in the proposed user shopping history enrichment model and compared the obtained results. Our study's focus in this section is closely tied to our second research question.

\item Another contribution of our study is the detailed analysis we made about how much enrichment we should do in the proposed shopping history enrichment approach. By measuring the direct effect of the number of imaginary masks we added to a user shopping history on performance, we have investigated an essential point for the correct application of the enrichment approach. This contribution enables us to answer research question 3.

\end{itemize}

We have organized our paper as follows: The introduction is covered in Section \ref{introduction}, while the related works are described in Section \ref{background}. Our proposed recommendation system is presented in Section \ref{methodology}, and the datasets, baseline models, implementation details, and evaluation results are discussed in Section \ref{results}. Finally, we present our conclusions in Section \ref{conclusions}.

\section{Related Works\label{background}}
Our review of literature on recommendation systems categorizes them into two main types: those based on traditional techniques and those employing deep learning techniques. The traditional recommendation systems are further divided into three categories \citep{AdomaviciusSurvey}: Collaborative Filtering, Content-Based, and Hybrid Recommendation Models. Collaborative Filtering models focus on user-item interaction to suggest the next item, whereas Content-Based models integrate item content into their recommendation logic. Hybrid models, combining both strategies, aim to leverage the strengths of each to enhance recommendation effectiveness.
Deep learning-based approaches, increasingly popular in recent years, are noted for their adeptness at modeling item relationships within user shopping histories. Initial deep learning studies, such as those using Multi-Layer Perceptron (MLP) \citep{MLP2, MLP3, MLP1}, demonstrated improved performance over earlier recommendation systems by learning the non-linear relationship between features. Further developments like DeepFM \citep{DeepFM} and Wide \& Deep Learning \citep{Wide&Deep} achieved better outcomes by combining linear and non-linear features. However, a common limitation in these approaches was their equal weighting of old and new user interactions, overlooking the evolving nature of user preferences.
Later studies aimed to address this by developing systems that account for the dynamic changes in user interests. These models, utilizing various sequential modeling techniques based on Recurrent Neural Networks (RNN), include Gated Recurrent Unit (GRU) based systems \citep{GRU1, Gru4Rec, GRU6, Gru4Rec+, GRU7, GRU2, GRU5, GRU4, GRU3} and Long-Short Term Memories (LSTM) based systems \citep{LSTM3, LSTM4, LSTM2, LSTM1, zhu2019learning}, treating user interest as a continually evolving process. A notable study employed horizontal and vertical convolutions for modeling user behavior \citep{Caser}.
The success of attention models in sequential modeling in the Natural Language Processing (NLP) domain inspired their application in recommendation systems \citep{CHEN202057, Attention2, Narm, li2020time, Attention4, Attention1, Wen2021, wu2019npa, Attention3}. The self-attention-based SASRec model \citep{SasRec} showed remarkable performance among these. Zhang et al. introduced an attention-based mutual structure for simultaneous predictions of the user's next favorite item and future audience \citep{zhang2021deep}. Recent studies have also adapted the bidirectional encoder representations model \cite{Bert} for recommendation, like BERT4Rec \cite{bert4rec}, which utilizes BERT's deep bidirectional self-attention capability to understand item relations in a user's shopping history. HybridBert4Rec \cite{hybridbert4rec} combines Content-Based and Collaborative Filtering with BERT, utilizing it for feature extraction and finding similar users. Additionally, there are studies \cite{islek2020hybrid, islek2022hierarchical, qiu2021u} that leverage BERT for content encoding.

Despite these advancements, recommendation models still grapple with data sparsity issues, as users often engage with multiple platforms, limiting the comprehensiveness of their shopping history on any single platform. Our approach, distinct from others, focuses on enriching the user's shopping history to provide more accurate recommendations. Additionally, we have adapted the bidirectional encoder representations model \cite{Bert}, initially developed for understanding word relationships in NLP, to model item relationships in shopping histories.

\section{The Proposed Recommendation System \label{methodology}}

\subsection{Problem Statement}

In the context of hierarchical recommendation system, we consider a group of users, represented as $\mathcal{U}=\{u_1, u_2, . . . , u_{|\mathcal{U}|}\}$, and a collection of items, denoted as $\mathcal{V} =\{v_1, v_2, . . . , v_{|\mathcal{V}|}\}$. For each user $u$, their interaction with items over time is recorded as a sequence $S_u=[v^{(u)}_1, v^{(u)}_2, ..., v^{(u)}_{n_u}]$, where each $v^{(u)}_t$ is an item from $\mathcal{V}$ engaged by user $u$ at a specific time $t$, and $n_u$ denotes the total number of interactions by user $u$. It is important to note that the length of this interaction sequence varies across users.

Based on this sequence $S_u$, our hierarchical recommendation system is designed to recommend the subsequent item $v^{(u)}_{n_u+1}$ that will likely capture the interest of user $u$.

\subsection{The Recommendation System Architecture}

\begin{figure}[h!]
  \centering
  \includegraphics[height=200px]{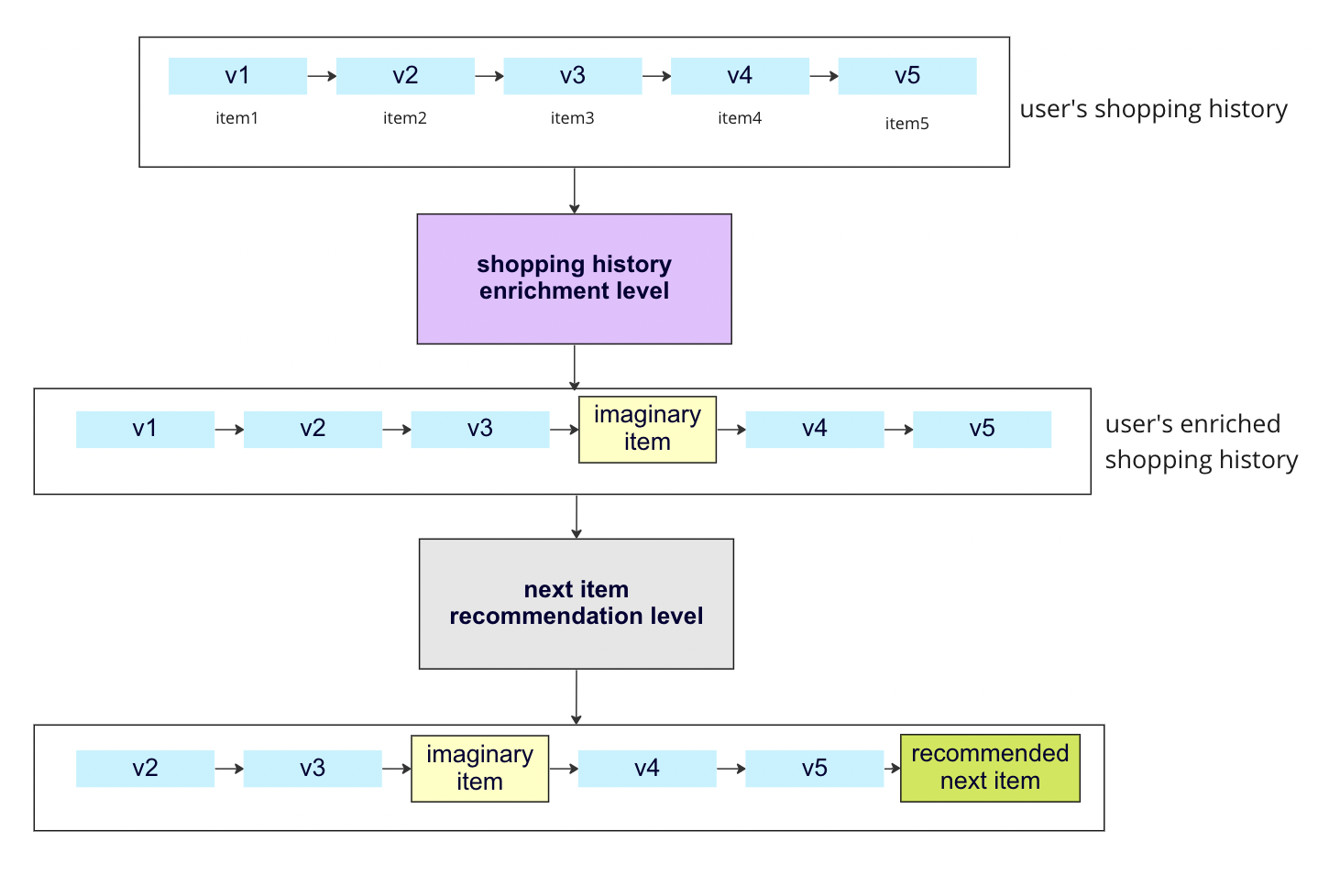}
  \caption{The Hierarchical Recommendation System.}
  \label{fig:basic_steps}
\end{figure}

We have developed a hierarchical item recommendation system for the e-commerce industry that takes into account the user's shopping history. The system is made up of two main levels, as shown in Fig. \ref{fig:basic_steps}: the shopping history enrichment level and the next item recommendation level. This approach aims to provide more personalized and accurate recommendations to users based on their shopping behavior.

\begin{figure}[h]
  \centering
  \includegraphics[scale=.4]{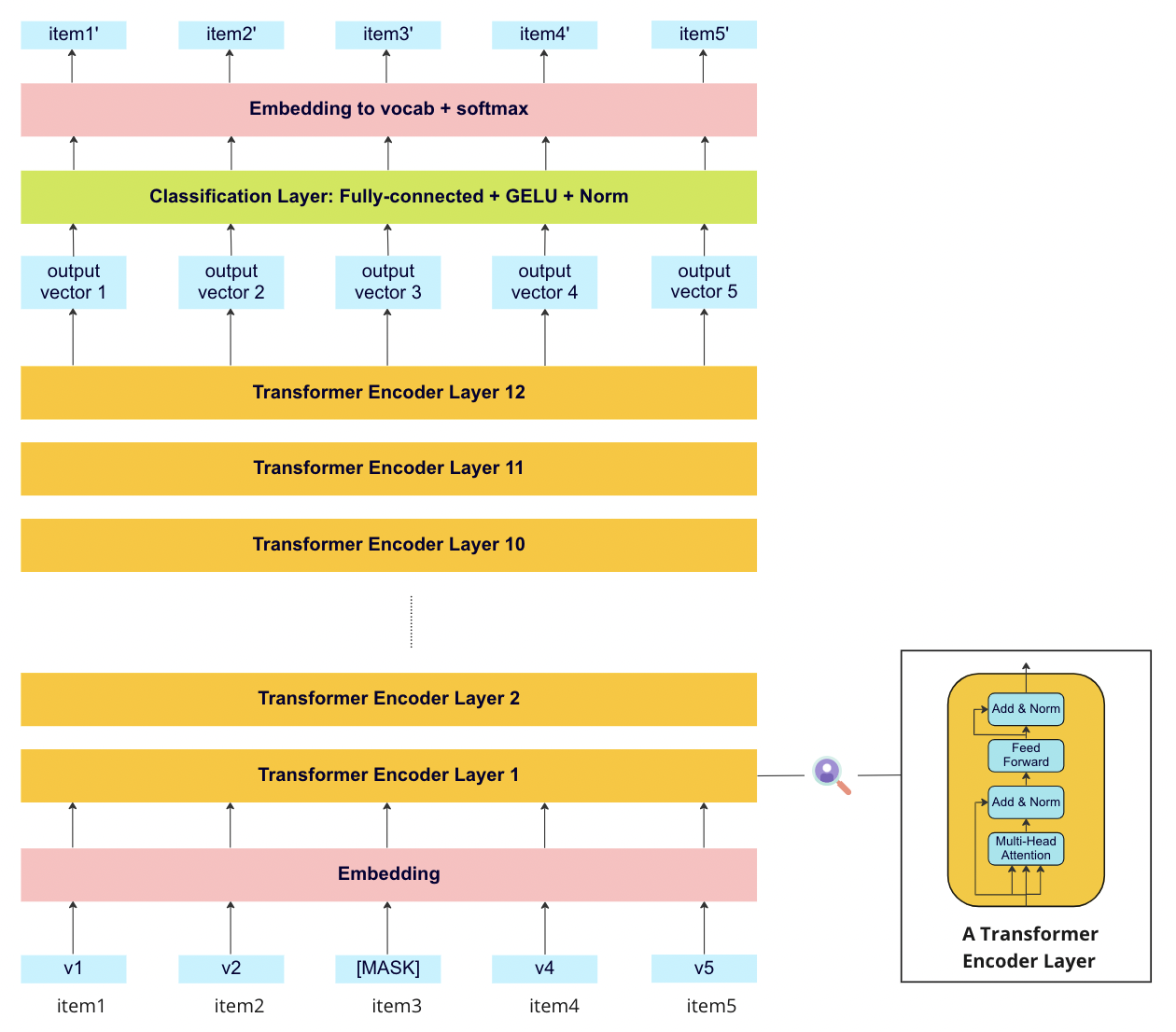}
  \caption{Bidirectional Encoder Representations Based Shopping History Enrichment Model.}
  \label{fig:bert}
\end{figure}

In the user's shopping history enrichment level, we take the shopping histories of the users $S_u$, which we separate for the training dataset, and we consider each user history as a sequence. We mask various items selected randomly in this sequence based on different scenarios with a probability of 15\%. Our approach involves training a bidirectional encoder representation model (BERT) \cite{Bert}. This architecture aims to predict which item is most probable to be found in a randomly masked position within a sequence of shopping history. The BERT model comprises 12 transformer layers, each with a length of 768, as illustrated in Fig.\ref{fig:bert}.

For each imaginary mask $m_i$, the bidirectional encoder representation-based history enrichment model predicts the most likely item to be placed for the selected imaginary mask position. The prediction for each imaginary mask is independent, and the initial shopping history is used during each prediction. In this way, a user's shopping history of length $n$ will have a length of $n+j$  as a result of this step, where $j$ is the number of added imaginary items to the user's shopping history.

As shown in Fig.\ref{fig:bert}, the Bidirectional Encoder Representations model contains several sequential Transformer Encoder Layers. Each Transformer Encoder Layer has a Multi-Head Attention Layer that takes a sequence as input. For instance, the first Transformer Encoder Layer takes the user's shopping history $S_u$ as input. In the Multi-Head Attention Layer, the Eq.\ref{eq:scaled_dot} is applied to every shopping history $S_u$ in the batch. In this Equation, $Q$, $K$, and $V$ are all equal to the chronologically ordered list of items in the user's shopping history sequence. 

In Eq.\ref{eq:scaled_dot}, $Q$ is a matrix that contains one row for each query. The shape of the $Q$ is [$n_{queries}$, $d_{keys}$] where $n_{queries}$ is the number of queries and $d_{keys}$ is the dimensions of each key. $K$ is another matrix that contains one row for each key. The shape of the $K$ is [$n_{keys}$, $d_{keys}$] where $n_{keys}$ is the number of keys. The number of values is also equal to the number of keys. $V$ is the matrix that contains one row for each value. The shape of the $V$ is the same as that of the $K$.

\begin{equation}
    \label{eq:scaled_dot}
    Attention(Q,K,V) = softmax\Big(\frac{QK^{T}}{\sqrt{d_{keys}}}\Big)V
\end{equation}

In Eq.\ref{eq:scaled_dot}, the $QK^{T}$ contains a similarity score for each query/key pair. The output is [$n_{queries}$, $n_{keys}$], but all rows sum up to 1. 
The score matrix is then scaled by the square root of the key vectors' dimension, which helps stabilize the training process by avoiding saturation. 
\begin{figure}[h!]
  \centering
  \includegraphics[height=180px]{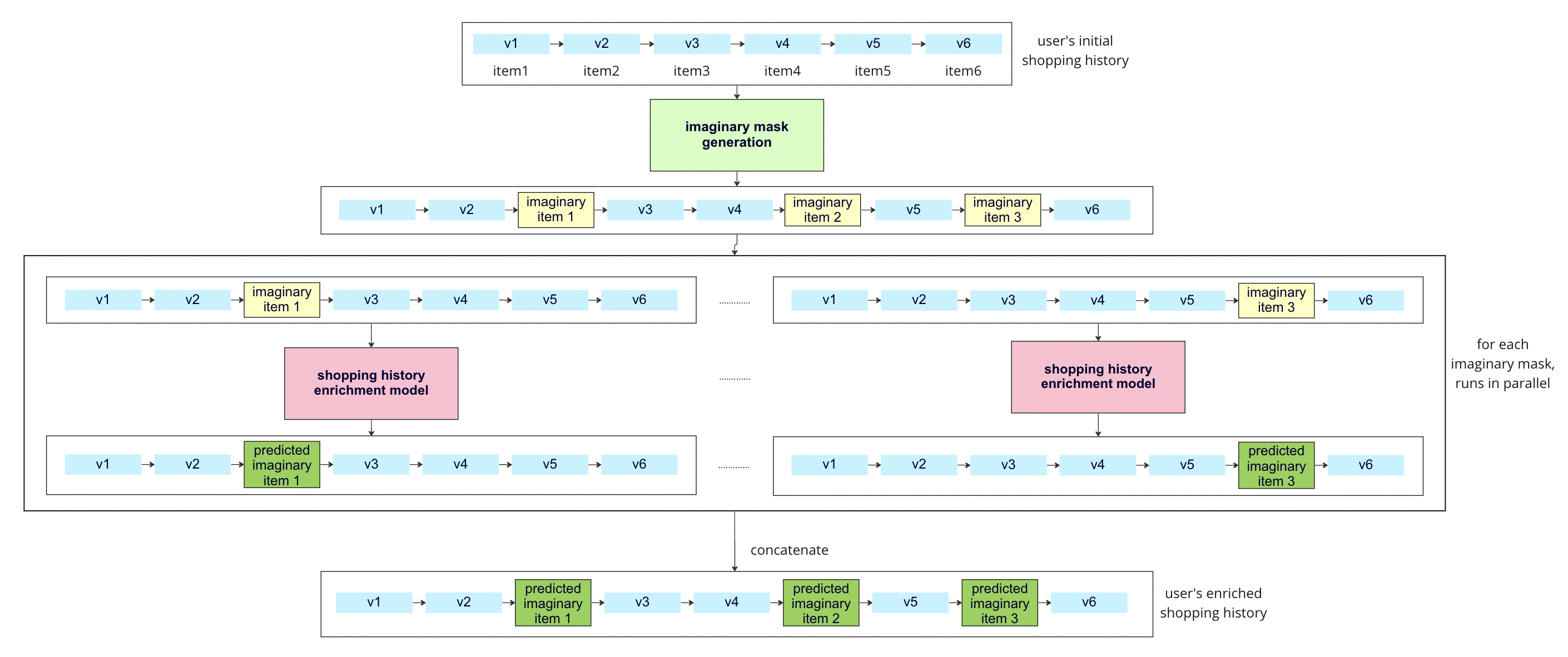}
  \caption{Shopping History Enrichment Level.}
  \label{fig:level1}
\end{figure}
After training the bidirectional encoder representation model to predict an intermediate item in a user's shopping history, we use this model to enrich the user's shopping history. As shown in Fig.\ref{fig:level1}, we place several imaginary masks between items in the user's shopping history sequence $S_u$ according to an imaginary mask generation approach. As seen from the Results section, we try two imaginary mask generation approaches: (1) randomly placed several imaginary masks in the shopping history according to the selected proportion, (2) placed imaginary masks between shopping sessions in the shopping history.

The next item recommendation model is the second layer of the hierarchical recommendation system. We use a self-attention sequential model called SASRec \citep{SasRec} as the next item recommendation model. The input of the sequential recommendation model is the enriched shopping history of user $u$. The enriched shopping history $E_u$ contains the ids of items purchased by user $u$ in chronological order.

\begin{figure}[h!]
  \centering
  \includegraphics[scale=.4]{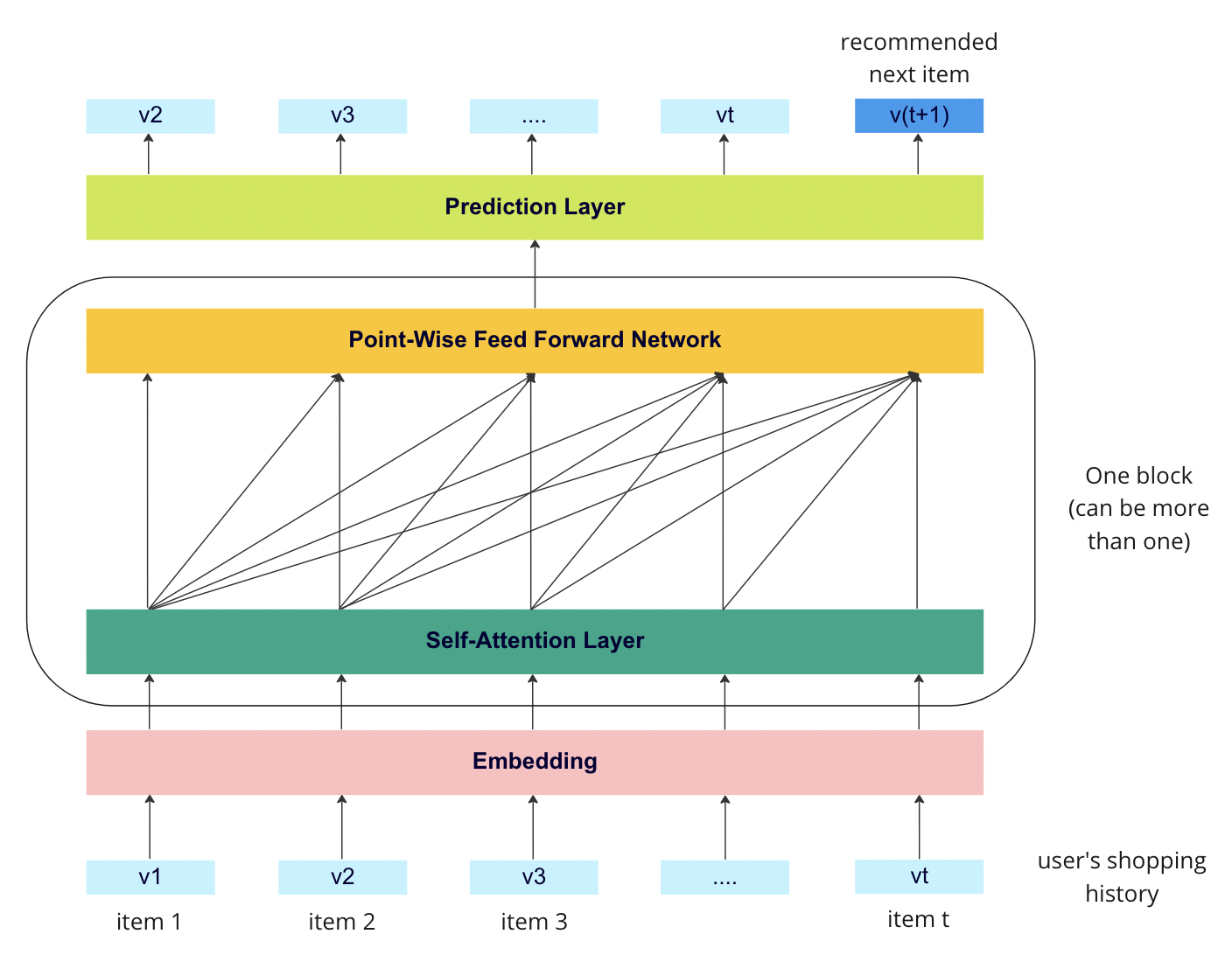}
  \caption{The Next Item Recommendation Model.}
  \label{fig:sasrec}
\end{figure}

In the embedding layer of Fig.\ref{fig:sasrec}, an embedding matrix $H_u$ of user $u$ is obtained from the enriched shopping history $E_u$. Item embedding vector $e^{(u)}_i$ and position embedding $p_i$ are generated for each item $v_i$ in the enriched shopping history $E_u$. The position embedding indicates the order of items in the user's shopping history. It means that a vector is assigned to represent every spot in the user's enriched shopping history, denoted as $E_u$.

\begin{equation}
    \label{eq:embedding_matrix}
    H_u = \begin{bmatrix} e^{(u)}_1 + p_1 \\ e^{(u)}_2 + p_2 \\ ... \\ e^{(u)}_{n_u} + p_{n_u}  \end{bmatrix}
\end{equation}

The diagram in Fig.\ref{fig:sasrec} illustrates that it is feasible to incorporate multiple self-attention blocks after the embedding layer in the model for recommending the next item in a sequence. Each self-attention block consists of a self-attention layer and a point-wise feed-forward network. The input for the $l$th block of self-attention, $H^l$, undergoes a transformation into three different matrices via linear projections $W^Q \in \mathbb{R}$, $W^K \in \mathbb{R}$, and $W^V \in \mathbb{R}$. These are trainable parameters that start with random initial values. The self-attention mechanism is detailed in Eq.\ref{eq:sequential_attention}.

\begin{equation}
    \label{eq:sequential_attention}
    S = SA({H^l}) = Attention(H^lW^Q, H^lW^K, H^lW^V)
\end{equation}

Equation \ref{eq:scale_dot_attention} illustrates the scaled-dot product attention mechanism. In this equation, $Q$ is used to denote queries, $K$ is employed to represent keys, and $V$ is for values. Precisely, the query corresponds to the input element $v_i$ in this context. The key, denoted as $K$, matches an item from the input shopping history. Meanwhile, $V$ stands for the value that reflects the interaction between the query element $Q$ and the key element $K$.

\begin{equation}
    \label{eq:scale_dot_attention}
    Attention(Q,K,V) = softmax\Big(\frac{QK^{T}}{\sqrt{d}}\Big)V
\end{equation}

The self-attention mechanism functions as a linear model that aggregates item embeddings from shopping history, utilizing trainable weights. Incorporating a point-wise feed-forward network into the self-attention layer's output significantly boosts non-linearity and accounts for the interactions across various latent dimensions. The configuration of the point-wise feed-forward network is outlined in Eq. \ref{eq:feed_fprward_network}. In this structure, $W^{(1)}$ and $W^{(2)}$ are indicative of the weight matrices, whereas $b^{(1)}$ and $b^{(2)}$ are indicative of the bias vectors.

\begin{equation}
    \label{eq:feed_fprward_network}
    F(H^l) = FFN(H^l) = RELU(H^lW^{(1)}+b^{(1)})W^{(2)} + b^{(2)} 
\end{equation}

Following the acquisition of relevant information through $l$ self-attention blocks, the prediction of the next item relies on $F$. In Eq.\ref{eq:relevance}, $r_{v, E_u}$ signifies the relevance score for item $v$ as the following item based on prior items $E_u$. The item embedding matrix $N \in \mathbb{R}^{|\mathcal{V}|xd}$ is introduced, where $d$ represents the embedding vector's length, and $|\mathcal{V}|$ denotes the size of the entire set of items.

\begin{equation}
    \label{eq:relevance}
    r_{v, E_u} = F(H^l)N^{T}
\end{equation}

The items are ranked based on relevance scores, with the most relevant item having the highest score for the user $u$.

This particular equation \ref{eq:binary_cross_entropy_loss} represents the output item of the recommendation model as $r_{v_{exp},t}$. Conversely, $r_{v_{neg},t}$ is used to symbolize a negative item chosen randomly at each time step $t$ from the shopping histories denoted by $E_u$.

\begin{equation}
    \label{eq:binary_cross_entropy_loss}
    \mathcal{L} = - \sum_{E_u \in S} \sum_{t \in \{1,...,n_u\}}[log(\sigma{(r_{v_{exp},t})}) + log(1 - \sigma{(r_{v_{neg},t})})]
\end{equation}

\section{Results \& Discussions\label{results}}
\subsection{Dataset\label{datasets}}
This research involves the evaluation of three separate datasets from Amazon, each corresponding to a different category: Movies \& TV, Video Games, and Beauty. Notably, these datasets do not include data on who bought an item. Hence, we considered each user instance reviewing an item as an "action" indicating their interest. Additionally, we have a text review for each action. However, we focused mainly on user-item actions and not on the meta-data of each item, such as item images, titles, and descriptions. To create a shopping history for each user, we sorted the items based on the time of user comments. Additionally, we removed the users and items that had fewer than five related actions. For more information on the selected datasets, please refer to Table \ref{tbl1}.

\begin{table*}[width=\linewidth,cols=5,pos=h]
\caption{Analyzes of datasets. }\label{tbl1}
\begin{tabular*}{\linewidth}{@{} LRRRRR@{} }
\toprule
Dataset name & \#users & \#items & \#actions &  avg. \#actions per user & avg. \#actions per item \\
\midrule
Beauty & 1.4k & 1.7k & 8.1k & 5.8 & 4.7 \\
Movies \& TV & 311k & 142k & 3.6m & 11.6 & 25.2 \\
Video Games &  64k  & 47k  & 585k & 9.1 & 12.3 \\
\bottomrule
\end{tabular*}
\end{table*}

\subsection{Evaluation Strategy \& Metrics\label{evaluation}}
In recommendation systems, two commonly used evaluation metrics are Hit Rate (HR) and Normalized Discounted Cumulative Gain (NDCG). We have decided to use these metrics to evaluate our work as it makes it easier to compare with the existing literature. 

\begin{itemize}
	\item \textbf{Hit Rate@k:}
 HR measures the percentage of test cases where the recommended item is included in the top-K recommended items list. It assesses the system's capability to recommend relevant items to the user. However, HR only considers the presence of relevant items in the recommendation list and not their ranking. The calculation of the HR@k metric can be observed in Equation \ref{eq:hitrate}.

    \begin{equation}
        \label{eq:hitrate}
        HR@k=\frac{|\{relevant\}\cap\{recommended@ k\}|}{|\{relevant\}|}
    \end{equation}

\item \textbf{Normalized Discounted Cumulative Gain:}
The calculation of Discounted Cumulative Gain (DCG) for a specified rank position $k$ is illustrated in Eq. \ref{eq:ndcg1}. In this formula, the variable $k$ is indicative of the rank at which relevance is calculated, and $relevant_{i}$ is a measure of the relevance level of the $i$th result.
\begin{equation}
\label{eq:ndcg1}
DCG_{k}=\sum_{i=1}^{k}\frac{relevant_{i}}{log_{2}(i+1)}
\end{equation}
For ranking items in a list according to their relevance, the IDCG value is computed as per Eq. \ref{eq:ndcg2}. Here, $R_k$ signifies the sequence of relevant items, sorted based on their relevance, up to the $k$th position in the dataset.
\begin{equation}
\label{eq:ndcg2}
IDCG_{k}=\sum_{i=1}^{|R_{k}|}\frac{2^{relevant_{i}} - 1}{log_{2}(i+1)}
\end{equation}
The normalized version of DCG, known as NDCG, is obtained by dividing the DCG score by the IDCG score, as shown in Eq.\ref{eq:ndcg3}.
\begin{equation}
\label{eq:ndcg3}
NDCG_{k}=\frac{DCG_{k}}{IDCG_{k}}
\end{equation}
The overall performance of the model is evaluated by averaging the NDCG scores across all rank positions.
\end{itemize}

For the shopping history enrichment model training, we use the user's shopping history without the last item. We randomly mask items within the remaining shopping history. For the next item recommendation model, we use the same training and test datasets of the shopping history enrichment model. We exclude the last item in the user's shopping history as the next item and use the remaining history as input and then train the next item recommendation model for predicting the last item.

As part of the evaluation process for our end-to-end hierarchical recommendation system, we randomly selected ninety-nine items that the user did not show attraction to. In the evaluation, the final item from the user's shopping history, along with ninety-nine other chosen negative items, form the test set. This set is utilized to assess the performances of models through the metrics of Hit-Rate@k and NCDG@k. For the purpose of our study, we have determined the value of 'k' to be 10.

Our approach is implemented using TensorFlow. For shopping history enrichment model, we use official BERT implementation. For the next item recommendation model, we use the official implementation of SASRec and the related hyperparameter set given in the paper \cite{SasRec}.

\subsection{Baselines\label{baselines}}
In this study, we compare our hierarchical recommendation model with the following existing baselines.     
\begin{itemize}
    \item \textbf{SASRec:}
    SASRec \citep{SasRec} serves as the baseline model and uses item ids for item representation.
    \item \textbf{DeepCoNN:}
    This model, DeepCoNN \citep{zheng2017joint}, uses two different TextCNN architectures to represent item and user reviews, respectively. The ratings model uses these latent factors learned from reviews.
    \item \textbf{NARRE:}
    This approach learns an attention weight and latent factors from reviews thanks to TextCNN \citep{chen2018neural}.
    \item \textbf{BERT4Rec:}
    A state-of-the-art model that utilizes bidirectional transformers for next item recommendations \citep{bert4rec}.    
    \item \textbf{SSE-PT:}
    The current research introduces the Personalized Transformer model (SSE-PT) \citep{ssept} to resolve the issue of insufficient personalization, as SASRec lacks inherent personalization and does not incorporate personalized user embeddings. 

 \end{itemize}   
The selected baseline models are state-of-the-art studies evaluated for different categories of related e-commerce datasets in the literature. Also, some baselines also utilize transformers. However, no study in the literature applies shopping history enrichment.

For the SASRec model \citep{SasRec}, the selected hyperparameters are as follows: a learning rate of 0.001, a batch size of 128, a dropout rate of 50\%, a maximum sequence length of 50, 50 hidden units, and 2 blocks.
Concerning the DeepCoNN model, as outlined by \citep{zheng2017joint} and \citep{sachdeva2020useful}, the hyperparameter configuration includes a convolutional kernel window of size 3, 100 neurons in the convolutional layer, a word embedding dimension of 300, a learning rate of 0.01, and a dropout ratio of 50\%.
For the NARRE model \citep{chen2018neural}, the hyperparameters are set with a convolutional kernel window size of 3, 100 neurons in the convolutional layer, an embedding dimension of 300, and a dropout ratio of 50\%.
In the TransNets model \citep{catherine2017transnets}, as per the study, the hyperparameters include a window size of 3 for the convolutional kernel, 100 neurons in the convolutional layer, an embedding dimension of 64, and a dropout ratio of 50\%. These settings were consistently used in all their experiments. With the BERT4Rec model \citep{bert4rec}, the learning rate is 0.0001, weight decay is 0.01, featuring two layers and two heads. In the case of SSE-PT \citep{ssept}, the model utilizes a learning rate of 0.001, an exponential decay rate for momentum, and a batch size of 128, comprising 2 encoder blocks. For all of our experiments, we have used the exact same hyperparameter settings that were originally reported by the authors.

The experiments and their results are detailed in Section \ref{test_results}.


\subsection{Evaluation Results \& Related Discussions\label{test_results}}

\begin{table*}[width=\linewidth,cols=5,pos=h]
\caption{End-to-End Test Scenarios.}\label{tbl2}
\begin{tabular*}{\linewidth}{cp{12cm}}
\toprule
End-to-end test identifier & \hfil Detail of the test scenario  \\
\midrule
End-to-end test scenario 1 & Predicting the next item when randomly masked items in the user's shopping history are removed from the sequence\\
End-to-end test scenario 2 & Predicting the next item using the user's shopping history directly\\
End-to-end test scenario 3 & Predicting the next item with the completed shopping history after added the randomly masked \%20 percent of items in the user's shopping history with the "shopping history enrichment model"\\
End-to-end test scenario 4 & Predicting the next item with the completed shopping history after added the randomly masked \%30 percent of items in the user's shopping history with the "shopping history enrichment model"\\
End-to-end test scenario 5 & Predicting the next item with the completed shopping history after added the randomly masked \%40 percent of items in the user's shopping history with the "shopping history enrichment model"\\
End-to-end test scenario 6 & Predicting the next item with the completed shopping history after added the randomly masked \%50 percent of items in the user's shopping history with the "shopping history enrichment model"\\
End-to-end test scenario 7 & Predicting the next item with the completed shopping history after added the randomly masked \%60 percent of items in the user's shopping history with the "shopping history enrichment model"\\
End-to-end test scenario 8 & Predicting the next item with the completed shopping history after added an imaginary masked item between different sessions in the user's shopping history with the "shopping history enrichment model"\\
End-to-end test scenario 9 & Predicting the next item with the completed shopping history after placing the top 2 items consecutively at that position for an imaginary masked item position between different sessions in the user's shopping history with the "shopping history enrichment model"\\
\bottomrule
\end{tabular*}
\end{table*}

Table \ref{tbl2} shows the test scenarios defined for the end-to-end detailed tests of the proposed hierarchical recommendation system. In "End-to-end test scenario 1", the $n$th item is predicted by the next item recommendation model without using randomly masked items ($n$ is the number of items in the shopping history, $m$ is the number of items that are masked from the input sequence) in the user's shopping history, that is, by using only $n-m-1$ items from the user's shopping history sequence. This test scenario is used for comparing the performances of the other scenarios, as a baseline. 

In "End-to-end test scenario 2", the items the first $n-1$ items in the user's shopping history are used directly for predicting $n$th item which is the last item in the shopping history.

In "End-to-end test scenario 3", approximately $(n-1)*0.2$ percent of imaginary masks are placed randomly in the user's shopping history according to the specified masking probability. 
The shopping history enrichment model predicts the most likely item for each imaginary mask. Afterward, the last item is predicted by the next item recommendation model using the resulting enriched shopping history, which has an average length of $(n-1)*0.2+(n-1)$.

The difference between "End-to-end test scenario 3" followed by 4 test scenarios is that the imaginary mask percentage that determines the number of imaginary masks is different. The imaginary mask percentages in the following 4 test scenarios are 0.3, 0.4, 0.5, and 0.6, respectively.

In "End-to-end test scenario 8", an imaginary mask is placed between different sessions in the shopping history. According to the timestamps of each user-item interaction, we define the shopping sessions of the users. We consider purchases made on different days are different shopping sessions. In this case, the total imaginary mask count equals $m$. The next item recommendation model predicts the next item uses the resulting enriched shopping history, which is $(n-1)+m$ length. 

In "End-to-end test scenario 9", an imaginary mask is placed between different sessions in the sequence. 
In this case, the total imaginary item count is equal to $m*2$ because the top two most likely items for an imaginary mask position are placed consecutively at that position between different sessions in the shopping history with the "shopping history enrichment model". The next item recommendation model predicts the last item using the enriched $(n-1)+m*2$ length shopping history.

The evaluation results of the end-to-end test scenarios are given in Table \ref{tbl3}. Each test scenario is run 10 times, and the results are given as the average of these 10 runs. Firstly, we provide an answer for RQ1 (Does enhancing the user's shopping history improve the recommendation model's performance?). In Table \ref{tbl3}, we assume "End-to-end test scenario 2" as a baseline, and we compare its results with "End-to-end test scenario 8", which we obtained our best results. As can be remembered, while in "End-to-end test scenario 2" we directly used the shopping history of the user, in "End-to-end test scenario 8", we placed imaginary masks between shopping sessions to enrich the shopping history of the user before recommending the next item. These results show that enriching the users' shopping history using imaginary masks improves recommendation performance for both NDCG@10 and HR@10 metrics.

\begin{table*}[width=\linewidth,cols=4,pos=h]
\caption{Next Item Prediction Evaluation Results. }\label{tbl3}
\begin{tabular*}{\linewidth}{@{} LLRR@{} }
\toprule
Dataset name & Test scenario & NDCG@10 & HR@10 \\
\midrule
\multirow{3}{*}{
Beauty} & Test scenario 1 & 0.5138 & 0.6474 \\
& Test scenario 2 & 0.5373 & 0.6725 \\
& Test scenario 3 & \textbf{0.5430} & 0.6793 \\
& Test scenario 4 & 0.5263 & 0.6735 \\
& Test scenario 5 & 0.5087 & 0.6723 \\
& Test scenario 6 & 0.4778 & 0.6392 \\
& Test scenario 7 & 0.4320 & 0.6070 \\
& Test scenario 8 & 0.5422 & \textbf{0.6821} \\
& Test scenario 9 & 0.4246 & 0.6053 \\
\midrule
\multirow{3}{*}{Movies \& TV} & Test scenario 1 & 0.5352 & 0.7791  \\
& Test scenario 2 & 0.5974 & 0.8135 \\
& Test scenario 3 & 0.6019 & 0.8367 \\
& Test scenario 4 & \textbf{0.6195} & 0.8475 \\
& Test scenario 5 & 0.5815 & 0.8217 \\
& Test scenario 6 & 0.5490 & 0.7851 \\
& Test scenario 7 & 0.5614 & 0.7825 \\
& Test scenario 8 & 0.6150 & \textbf{0.8526} \\
& Test scenario 9 & 0.5798 & 0.8094 \\
\midrule
\multirow{3}{*}{Video Games} & Test scenario 1 & 0.4406 & 0.6990 \\
& Test scenario 2 & 0.5047 & 0.7825 \\
& Test scenario 3 & 0.5171 & 0.8009 \\
& Test scenario 4 & 0.5204 & 0.8177 \\
& Test scenario 5 & 0.5228 & 0.8216 \\
& Test scenario 6 & 0.4839 & 0.7564 \\
& Test scenario 7 & 0.4635 & 0.7408 \\
& Test scenario 8 & \textbf{0.5296} & \textbf{0.8325} \\
& Test scenario 9 & 0.4930 & 0.5349 \\
\bottomrule
\end{tabular*}
\end{table*}

For RQ2 (Which of the different approaches to determining which location to choose for enrichment in the user's shopping history provides more precise user shopping history enrichment?), should be compared the performances of the test scenarios 3 to 9. For the HR@10 metric, it is obvious for all three datasets that instead of randomly selecting the mask positions, placing them between shopping sessions gives the best results. Also, for NDCG@10 metric, placing imaginary masks between shopping sessions gives best results in Amazon Video Games dataset. For Amazon Beauty and Amazon Movies \& TV datasets, placing between sessions gives the second best results but there are little to no difference between best and second best results. 

Table \ref{tbl4} shows that "test scenario 8" had the least number of imaginary masks added, but yielded the best results. Here it can be concluded that improving the user's shopping history by adding a sufficient number of imaginary masks placed in the correct positions is the most appropriate approach. Also, this situation could be analyzed as: placing imaginary masks between shopping sessions where the user is more likely to have made a purchase from another e-commerce platform rather than choosing random positions, which is a more realistic approach. Another point is that considering a real-world e-commerce platform, the fewer imaginary masks there are, the fewer imaginary items will be predicted, so it is possible to improve recommendation system performance both in less time and with fewer transactions. 

\begin{table*}[width=\linewidth,cols=4,pos=h]
\caption{Imaginary Mask Analysis. }\label{tbl4}
\begin{tabular*}{\linewidth}{@{} LLRR@{} }
\toprule
Dataset name & Test scenario & Median Imaginary Mask Count &  Candidate Imaginary Mask Slot\\
\midrule
\multirow{3}{*}{Beauty} & Test scenario 3 & 1339 & 6711 \\
& Test scenario 4 & 1946 & 6711 \\
& Test scenario 5 & 2805 & 6711 \\
& Test scenario 6 & 3288 & 6711 \\
& Test scenario 7 & 4093 & 6711 \\
& Test scenario 8 & 336 & 6711 \\
& Test scenario 9 & 672 & 6711 \\
\midrule
\multirow{3}{*}{Movies \& TV} & Test scenario 3 & 309097 & 1561098 \\
& Test scenario 4 & 474572 & 1561098 \\
& Test scenario 5 & 624437 & 1561098 \\
& Test scenario 6 & 780520 & 1561098 \\
& Test scenario 7 & 952268 & 1561098 \\
& Test scenario 8 & 97568 & 1561098 \\
& Test scenario 9 & 195136 & 1561098 \\
\midrule
\multirow{3}{*}{Video Games} & Test scenario 3 & 96138 & 457800 \\
& Test scenario 4 & 143749 & 457800 \\
& Test scenario 5 & 182204 & 457800 \\
& Test scenario 6 & 232592 & 457800 \\
& Test scenario 7 & 274681 & 457800 \\
& Test scenario 8 & 15869 & 457800 \\
& Test scenario 9 & 31738 & 457800 \\
\bottomrule
\end{tabular*}
\end{table*}

\begin{figure}
  \centering
  \includegraphics[scale=0.35]{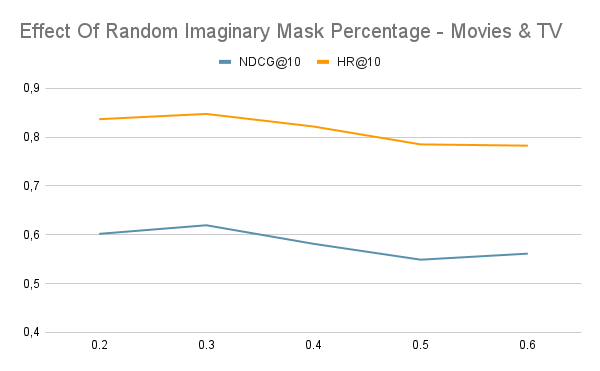}
  \caption{Effect Of Random Imaginary Mask Percentage - Movies \& TV.}
  \label{fig:mask_movies_tv}
\end{figure}

\begin{figure}
  \centering
  \includegraphics[scale=0.35]{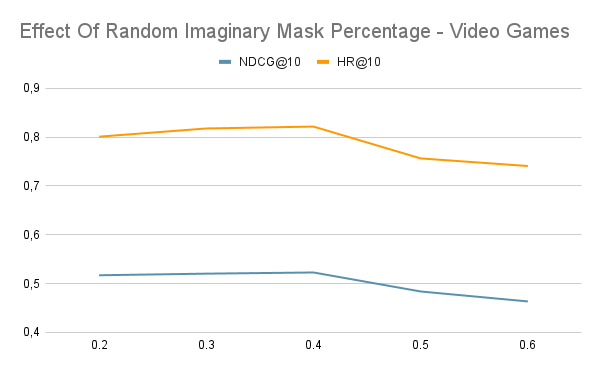}
  \caption{Effect Of Random Imaginary Mask Percentage - Video Games.}
  \label{fig:mask_video_games}
\end{figure}

The analysis for the RQ3 (When enriching the user’s shopping history, does the performance of recommending the next item increase as we continue to enrich?) can be seen in Fig. \ref{fig:mask_movies_tv} and Fig. \ref{fig:mask_video_games}.
The imaginary masks and the next item recommendation performance relation shows that the performance does not improve as the number of imaginary masks increases. Here, we analyze the approach of placing imaginary masks at random locations, and we see that the ratio of imaginary masks with the best results is different for each dataset. Here is our understanding: it is more important to put the imaginary masks in the right places than the number of them. If imaginary masks are put in the "right" position, there is a direct improvement in performance. However, as the number of "right" positions is limited, we observed no improvement in performance as the number of imaginary masks increased.

\begin{figure}
  \centering
  \includegraphics[scale=0.35]{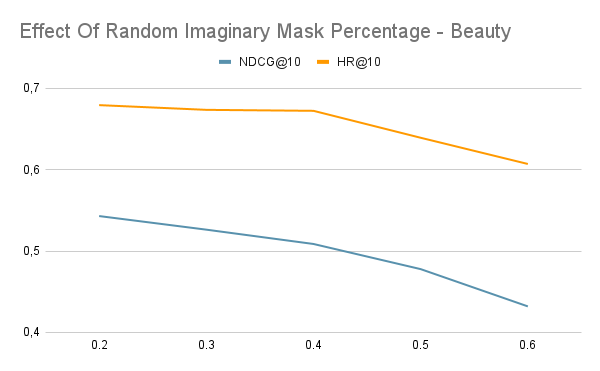}
  \caption{Effect Of Random Imaginary Mask Percentage - Beauty.}
  \label{fig:mask_beauty}
\end{figure}

The results for HR@10 and NDCG@10 for all datasets of models are presented in Table \ref{tbl5}. According to our knowledge, no study in the literature uses the shopping history enrichment approach, for this reason, we compare our study with default SASRec and other state-of-the-art recommendation studies. For Amazon Movies \& TV and Amazon Video Games datasets, we can say that the proposed shopping history enrichment-based recommendation system provides the best results in both NDCG@10 and HR@10 metrics. For the Amazon Beauty dataset, the proposed approach also gives the best HR@10 metrics. The reason we cannot obtain the best NDCG@10 results in Amazon Beauty could be that the items in Amazon Beauty dataset are quite similar rather than other datasets. Since it is a relatively limited category, the items it includes are very similar items of different brands. For example, for the black mascara item, many different brands have similar items.


\begin{table*}[width=\linewidth,cols=5, pos=h!]
\caption{Test Results Of Baseline Models. }\label{tbl5}
\begin{tabular*}{\linewidth}{@{} LLRR@{} }
\toprule
Dataset name & Model Name & NDCG@10 & HR@10  \\
\midrule
Beauty & SASRec & \textbf{0.5469} & 0.6610 \\
Beauty & DeepCoNN & 0.5178 & 0.6418 \\
Beauty & NARRE & 0.5206 & 0.6302 \\
Beauty & BERT4Rec & 0.5418 & 0.6778 \\
Beauty & SSE-PT & 0.5370 & 0.6736 \\
Beauty & Shopping History Enrichment Based RecSys & 0.5422 & \textbf{0.6821} \\
\midrule
Movies\&TV & SASRec & 0.5895 & 0.8144 \\
Movies\&TV & DeepCoNN & 0.3904 & 0.5980 \\
Movies\&TV & NARRE & 0.3953 & 0.6097 \\
Movies\&TV & BERT4Rec & 0.5876 & 0.8312 \\
Movies\&TV & SSE-PT & 0.5942 & 0.8388 \\
Movies\&TV & Shopping History Enrichment Based RecSys & \textbf{0.6150} & \textbf{0.8526} \\
\midrule
Video Games & SASRec & 0.4607 & 0.7091 \\
Video Games & DeepCoNN & 0.3228 & 0.4936 \\
Video Games & NARRE & 0.3386 & 0.5103 \\
Video Games & BERT4Rec & 0.5089 & 0.7520 \\
Video Games & SSE-PT & 0.5260 & 0.7750 \\
Video Games & Shopping History Enrichment Based RecSys & \textbf{0.5296} & \textbf{0.8325} \\
\bottomrule
\end{tabular*}
\end{table*}


A general review of the proposed hierarchic recommendation system is related to the usage scenario in real-world e-commerce platforms. With the scheduled jobs to be worked every night, the shopping history of each user should be enriched automatically. As soon as any user visits the e-commerce platform, the item that she/he will be interested in at that moment can be recommended by using the enriched and recorded shopping history of the user.

\section{Conclusions \label{conclusions}}

Our paper presents a hierarchical model that enhances users' shopping history. Real-world users shop from multiple e-commerce platforms simultaneously; in other words, different parts of the user's shopping history are shared between different e-commerce platforms. In light of this information, we assume in this study that no e-commerce platform has a complete record of the user's shopping history, but they can only access some parts of it. If the recommendation system predicts the missing parts and can enrich the user's shopping history properly, it will be possible to recommend the next item more accurately. 

There are two main levels of the proposed hierarchical recommendation system: In the first level, we enrich the user's shopping history thanks to a bidirectional encoder representations model. In the second level of the system, we recommend next items accurately. The proposed approach improves HR@10 and NDCG@10 metrics according to the obtained results. In other words, this hierarchical structure provides a recommendation system with higher performance for real-world e-commerce platforms.

As a future work, we will apply the proposed approach to the cross-domain recommendation problem. In this way, we think it is possible to increase cross-domain recommendations in the e-commerce domain.

\printcredits

\bibliographystyle{cas-model2-names}

\bibliography{cas-refs}


\end{document}